\documentclass[aps,showpacs]{revtex4}

\newcommand{\leftsuperindex}[2]{  { {}^{\mbox{\tiny{$(#2)$}}} \! #1 } }
\newcommand{\three}[1]{ \leftsuperindex{#1}{3} }
\newcommand{\four}[1]{ \leftsuperindex{#1}{4} }

\begin{document}


\title{Hamiltonian theory for the axial perturbations of a dynamical
spherical background}

\author{David Brizuela}
\author{Jos\'e M. Mart\'{\i}n-Garc\'{\i}a\footnote{
Present address:
Laboratoire Univers et Th\'eories, CNRS,
5 place Jules Janssen, F-92190 Meudon, FRANCE,
and \\
Institut d'Astrophysique de Paris, CNRS,
98bis bd Arago, F-75014 Paris, FRANCE
}
}
\affiliation{Instituto de Estructura de la Materia, CSIC, Serrano
121-123, 28006 Madrid, Spain}

\date{\today}


\begin{abstract}
We develop the Hamiltonian theory of axial perturbations around a general
time-dependent spherical background spacetime. Using the fact that
the linearized constraints are gauge generators, we isolate the physical
and unconstrained axial gravitational wave in a Hamiltonian pair of
variables. Then, switching to a more geometrical description of the
system, we construct the only scalar combination of them.
We obtain the well-known Gerlach and Sengupta scalar for axial perturbations,
with no known equivalent for polar perturbations. The strategy suggested
and tested here will be applied to the polar case in a separate
article.
\end{abstract}


\maketitle


\section{Introduction and overview}

Perturbation Theory (both linear and higher order) is one of the most
successful tools in General Relativity (GR). It has been used to find the
stability properties of a large variety of background solutions like
black holes, critical or cosmological solutions. It is also useful
to model the evolution of dynamical processes in astrophysical scenarios
that slightly deviate from an exact symmetry, like the oscillations
of a static spherical neutron star or a nearly-spherical supernova
explosion. In particular, it can be used to investigate the emission of
gravitational waves in those processes.

A central problem in GR perturbation theory, inherited from the
diffeomorphism invariance of the full theory, is that of isolating
the physical degrees of freedom from the gauge-dependent information
\cite{BMM07}.
This can be done by imposing convenient {\em gauge fixing} conditions
on the perturbations, as Regge and Wheeler \cite{ReWh57} originally
did in their study of perturbations of a Schwarzschild black hole.
They and later Zerilli \cite{Zeri70} succeeded in isolating the two
physical degrees of freedom of the gravitational field around spherical
vacuum, by taking suitable linear combinations of the remaining
perturbations and their radial derivatives. These two variables further
decouple due to their different properties under parity inversion:
the Regge-Wheeler variable is axial and the Zerilli variable is polar.

A more systematic treatment of the gauge freedom in GR perturbation
theory was pioneered by Moncrief \cite{Monc74} in his Hamiltonian study
of the nonspherical perturbations of Schwarzschild. In a Hamiltonian 
context the four constraints obeyed by the twelve dynamical gravitational
variables are the generators of the gauge transformations. Moncrief was
able to use this information to perform several canonical transformations
which reorganized the original six canonical pairs of variables into
two physical pairs (equivalent to the Regge-Wheeler and Zerilli variables
and their canonical momenta) and four gauge pairs in which the momenta
were constrained to be zero, without any gauge fixing. The same technique
was later applied to other spherical backgrounds with additional
symmetries, like
Reissner-Nordstr\"om \cite{Monc74b,Monc75}, Oppenheimer-Snyder \cite{CPM78}
or Friedmann-Robertson-Walker
\cite{Gund93}, but has never been applied to general spherically symmetric
backgrounds, possibly highly time dependent. Hamiltonian perturbation
theory has also been recently revisited in Quantum Gravity with a
cosmological background \cite{GHT07}. A drawback of the Hamiltonian
approach is that it is tied to a particular foliation of the background
spacetime, and hence the geometric properties of the gauge-invariant
variables under coordinate transformations involving time are far from obvious.

A {\em Lagrangian} formalism was introduced by Gerlach and Sengupta \cite{GeSe79}
(GS) to study perturbations around generic spherical spacetimes, in
which the metric perturbation is geometrically split along the
decomposition of the 4d manifold $M^4$ into the product of a general
2d Lorentzian manifold $M^2$ with boundary and the unit 2-sphere $S^2$.
This is a highly geometrical framework, in which the meaning of the
perturbations is transparent, and which also allows the construction
of gauge-invariant variables. In the axial case it has been possible
to isolate the gravitational degree of freedom in a single scalar
{\em master} variable which obeys a wave equation and can be coupled
to any kind of matter, both in the background and the perturbations. This
master scalar and its equation generalize the Regge-Wheeler variable and
its equation to the axial perturbative problem around spherical symmetry
for any reasonable matter model, and hence can be considered as the optimal
framework for a perturbative study. Unfortunately, in the polar case
there is not a master scalar valid for a generic spherical background and
any matter model, though there are results for some particular cases.
For instance, a master Zerilli scalar has been introduced
by Sarbach and Tiglio \cite{SaTi00} for a Schwarzschild background,
which was later generalized to nonlinear electrodynamics \cite{MoSa03},
around any background solution of the theory.
In references \cite{MaPo05,NaRe05} the gauge-invariant combinations of the
stress-energy tensor were also included but still on a vacuum background.

Both approaches to metric perturbation theory are complementary: the
Hamiltonian approach offers a better framework to handle gauge-invariance,
while the Lagrangian approach gives a clearer picture of the geometrical
structures being perturbed. This Article proposes a combination of both
formalisms to construct gauge-invariant scalar perturbative variables
containing all physically relevant information concerning the gravitational
waves. We restrict ourselves
to spherical backgrounds, but which can be highly dynamical. For
definiteness, the dynamics will be introduced using a real massless
scalar field, but could be done through any other matter model admitting
a Hamiltonian description. This Article focuses on the axial subset of
perturbations, for which the sought solution is the Gerlach and Sengupta
master scalar \cite{GeSe79}, previously found using the Lagrangian
method only.
We show how the Hamiltonian way allows a more systematic derivation of
this object, and how both approaches mutually relate. Most important, this
Article prepares the path towards a systematic analysis of the polar
problem, for which a general gauge-invariant master scalar has never
been found. Such analysis will be discussed in a second publication.


The Article is organized as follows. Section \ref{section:idea} presents
a brief review of the ADM formalism, establishing the notations for
3d and 4d objects, both in the background and first-order perturbations.
Section \ref{section:bg} particularizes to a spherical background.
Section \ref{section:axial} restricts to axial perturbations and
carries out the complete program of scalar gauge-invariant construction,
as well as establishing the connection between the Hamiltonian and
Lagrangian approaches to the problem.
We conclude in Section \ref{section:conclusions} with some remarks.

\section{Hamiltonian perturbations in General Relativity}
\label{section:idea}

\subsection{ADM Hamiltonian formalism}

Given the four-dimensional spacetime $(M^4,\four{g}_{\mu\nu})$,
we introduce a foliation of 3d spacelike slices as level surfaces of
the time field $t(x)$. The orthogonal vector $u^\mu$ defines
the projected metric $\three{g}_{\mu\nu}=\four{g}_{\mu\nu}+u_\mu u_\nu$
on the slices. We introduce coordinates $(t,x^i)$ adapted to the
foliation, and work with three-dimensional objects. Only in the last
part of this Article we shall use four-dimensional metric variables
in order to compare our results with those from the Gerlach and
Sengupta formalism.
Greek and Latin indices denote 4d and 3d tensors respectively.
A left-superindex indicates dimensionality when confusion may arise.

The 4-metric is decomposed as customary in the lapse function, the shift 
vector and the 3-metric on the slices
\begin{equation} \label{lapse&shift}
\alpha^{-2}\equiv - \four{g}^{tt} , \qquad
\beta_i \equiv \four{g}_{ti} , \qquad
g_{ij}\equiv\four{g}_{ij} ,
\end{equation}
with inverse
\begin{equation}
g^{ij}=\four{g}^{ij}+\alpha^{-2}\beta^i\beta^j ,
\end{equation}
with Latin indices always raised and lowered with $g^{ij}$ and $g_{ij}$.

The gravitational dynamical variables in the ADM Hamiltonian formalism 
are $g_{ij}$ and their conjugated momenta:
\begin{equation} \label{Pidef}
\Pi^{ij}\equiv\mu_g\left(g^{ij}K^l{}_l-K^{ij}\right) , \qquad
\mu_g\equiv\sqrt{\det g_{ij}} , 
\end{equation}
where $K^{ij}$ is the extrinsic curvature of the foliation hypersurfaces.

The spacetime will be assumed to contain a dynamical Klein-Gordon field
$\Phi$, whose evolution is controlled by the action
\begin{eqnarray}
{\cal S}_{KG} &=& - \frac{1}{2} \int dx^4 \sqrt{-\four{g}} \; 
                               \four{g}^{\mu\nu}
                               \Phi_{,\mu} \Phi_{,\nu} \\
&=& \int dx^4 \left[
\Pi \Phi_{,t} 
-\frac{\alpha}{2} \left(\frac{\Pi^2}{\mu_g}
             +\mu_g g^{ij}\Phi_{,i}\Phi_{,j}\right)
-\beta^i \left(\Pi \Phi_{,i} \right)
\right] .
\end{eqnarray}
Its canonical momentum has been defined as
\begin{equation}
\Pi\equiv-\sqrt{-\four{g}}\;\four{g}^{t\mu}\Phi_{,\mu} ,
\end{equation}
and $\four{g}$ denotes the determinant of the 4-metric.
The complete action of the system, with coupling constant
$16\pi G_N =1$ following \cite{Monc74}, is given by
\begin{equation} \label{action}
{\cal S} = {\cal S}_G + {\cal S}_{KG} =
\int dt \int d^3x \left( 
\Pi^{ij} g_{ij,t} 
+ \Pi \Phi_{,t}
- \alpha {\cal H} 
- \beta^i {\cal H}_i \right).
\end{equation}
The Lagrange multipliers
$\alpha$ and $\beta^i$ are associated to the constraints
\begin{eqnarray}
{\cal H} &=& \frac{1}{\mu_g}
\left[\Pi^{ij}\Pi_{ij}-\frac{1}{2}\left(\Pi^l{}_l\right)^2\right]
     - \mu_g \three{R} 
     + \frac{1}{2} \left( \frac{\Pi^2}{\mu_g}
                  +  \mu_g g^{ij} \Phi_{,i}\Phi_{,j}
\right), \\
{\cal H}_i &=&  -2 D_j\Pi_i{}^j + \Pi \Phi_{,i},
\end{eqnarray}
where $D_j$ is the covariant derivative associated to $g_{ij}$.
Variation of the action (\ref{action}) with respect to
$g_{ij}$, $\Pi_{ij}$, $\Phi$ and $\Pi$ gives the evolution equations
for the corresponding conjugated variables.

\subsection{Hamiltonian metric perturbations}
Now suppose that the whole system is perturbed at first order.
We define the following special notations:
\begin{eqnarray}
C \equiv \delta \alpha , \quad &\qquad&
B^i \equiv \delta(\beta^i) , \\
h_{ij} \equiv \delta(g_{ij}) , &\qquad& 
p^{ij} \equiv \delta(\Pi^{ij}) , \\
\varphi \equiv \delta \Phi , \quad &\qquad&
\quad p \equiv \delta \Pi .
\end{eqnarray}
Notice that we perturb the contravariant components of the shift
vector because this will give rise to simpler equations, even though
the comparison with GS variables will be slightly more involved because
$\beta_i$ is better related to the \mbox{4-metric}
[see Eq. (\ref{lapse&shift})].

Following Taub \cite{Taub} and Moncrief \cite{Monc74} we shall obtain
the equations for the linear perturbations using the Jacobi method of
second variations. The idea is that the second variation
of the action (\ref{action}), keeping only terms that are quadratic on
first-order perturbations, gives an action functional for the
perturbations,
\begin{equation}\label{effaction}
\frac{1}{2}\delta^2{\cal S} =
\int dx^4 \left[ 
p^{ij} h_{ij,t} 
+ p \varphi_{,t}
- C \delta({\cal H})
- B^i \delta({\cal H}_i)
-\frac{\alpha}{2} \delta^2({\cal H})
- \frac{\beta^i}{2} \delta^2({\cal H}_i)
\right] .
\end{equation}
There are three kinds of terms. First we have kinetic terms, containing
time derivatives of $h_{ij}$ and $\varphi$. Then we have the first
variations of the constraints that, under a variation of the effective
action (\ref{effaction}) with respect to $B^i$ and $C$, give the
constraints that must be obeyed by the perturbations,
\begin{equation}\label{constrainteq}
\delta({\cal H})=0, \qquad \delta({\cal H}_i)=0.
\end{equation}
And finally we have the second variations of the constraints,
which are quadratic in the perturbations $(h_{ij}, p^{ij},
\varphi, p)$, and will give the evolution equations for those
perturbations.
Even though we started with an exact Hamiltonian which was a linear
combination of constraints, we end up having a quadratic Hamiltonian
which does not vanish on shell.

The constraints ${\cal H}$ and ${\cal H}_i$ in General Relativity are
first-class constraints, and hence generators of gauge transformations
on the constraint surface in phase space. This identifies the gauge
orbits, but in general it is not possible to separate explicitly
the 2 physical degrees of freedom (4 functions) from the 4 gauge
variables and the 4 constrained variables in $g_{ij}$ and $\Pi^{ij}$.
The situation in
the linearized theory is simpler, but still only highly symmetric
background scenarios allow the construction of {\em gauge-invariant}
algebraic combinations of perturbations and their derivatives
containing the physical information in the linearized approximation.
(See \cite{BMM07} for a discussion of the importance of symmetry
on the algebraic character of the combinations.) One of such background
scenarios is a spherically symmetric spacetime, as we shall exploit
for the rest of this Article.


For completeness, we provide the expressions for the first variations of
the constraints:
\begin{eqnarray}
\delta({\cal H}) &=& 
\frac{1}{\mu_g}\left(\Pi_{ij}-\frac{1}{2}g_{ij}\Pi^l{}_l\right)
\left(2p^{ij}+2h^i{}_k\Pi^{kj}-\frac{1}{2}h^k{}_k\Pi^{ij}\right) \\
&+&\mu_g \left(\three{G}^{ij}h_{ij}-D^iD^jh_{ij}+D^jD_jh^i{}_i\right) 
\nonumber \\
&+& \frac{1}{4}h^k{}_k \left(-\frac{\Pi^2}{\mu_g}+\mu_g \Phi_{,i}\Phi^{,i}\right)
+\frac{p\Pi}{\mu_g}
+\mu_g\varphi_{,i}\Phi^{,i}
-\frac{1}{2}\mu_g h^{ij}\Phi_{,i}\Phi_{,j}
, \label{deltaH} 
\nonumber \\
\delta({\cal H}_i) &=& -2D_k (h_{ij}\Pi^{jk}+g_{ij}p^{jk})
+\Pi^{jl}D_ih_{jl}
+p \Phi_{,i}+\Pi\varphi_{,i}.
\label{deltaHi}
\end{eqnarray}
See \cite{BMM06} for intermediate expressions and techniques to compute
these expressions. The corresponding expressions for their
second variations are
\begin{eqnarray}
\delta^2({\cal H})&=&\frac{1}{8\mu_g}\{
8 p^2+8\mu_g^2\leftsuperindex{G}{3}^{ij}(h_{ij} h_k{}^k
-2 h_i{}^k h_{jk})+ 16 P_{ij} P^{ij}-8 P_i{}^i P_j{}^j
\\\nonumber
&+&2 h^{ij}\left[h^{kl}(8\Pi_{ik}\Pi_{jl}-4\Pi_{ij}\Pi_{kl})
+h_{ij}\left(\Pi^2-2 \mu_g^2\leftsuperindex{R}{3} +2\Pi_{kl}\Pi^{kl}
-\Pi_k{}^k\Pi_l{}^l\right)
\right.\\\nonumber
&-&\left.
8 \mu_g^2(D_jD_ih_k{}^k-D_jD_kh_i{}^k-D_kD_jh_i{}^k+D_kD^kh_{ij})
-8\Pi_k{}^k P_{ij}+32\Pi_i{}^k P_{jk}-8\Pi_{ij}P_k{}^k\right]
\\\nonumber
&+&h_i{}^i \left[h_j{}^j \left(\Pi^2+2\Pi_{kl}\Pi^{kl}
-\Pi_k{}^k \Pi_l{}^l+2\mu_g^2\leftsuperindex{R}{3}\right)
+8 h^{jk}\left( \Pi_{jk}\Pi_l{}^l -2 \Pi_j{}^l\Pi_{kl}
\right)
\right.\\\nonumber
&-&\left.8 \mu_g^2(D_kD_jh^{jk}-D_kD^kh_j{}^j)
-8\Pi p-16 \Pi^{jk} P_{jk}+8\Pi_j{}^jP_k{}^k\right]
\\\nonumber
&+&\mu_g^2 [8 D_i\varphi D^i\varphi+ h_j{}^jD_i\Phi
(8 D^i\varphi+h_k{}^k D^i\Phi)+2 h_{jk}D^i\Phi
(4 h_i{}^k D^j\Phi-h^{jk}D_i\Phi)
\\\nonumber
&+&4D_jh_k{}^k D^jh_i{}^i
-4 h_{ij} (4 D^i\varphi+h_k{}^k D^i\Phi)D^j\Phi
+16(D_ih^{ij}-D^jh_i{}^i)D_kh_j{}^k
+4(2D_jh_{ik}-3 D_k h_{ij})
D^kh^{ij}]
\},\\
\delta^2({\cal H}_i)&=& 2 p \varphi_{,i}+2 p^{jk}(D_ih_{jk}
-2D_k h_{ij})-4 h_{ij} D_k p^{jk}.
\end{eqnarray}

\section{Spherical background}
\label{section:bg}

Let us know restrict to a spherically symmetry background
$M^4=M^2\times S^2$, where $S^2$ is the unit two-sphere and $M^2$ is a
two-dimensional Lorentzian manifold with boundary.
We shall use arbitrary coordinates $x^A=(t,\rho)$ on $M^2$ and the
usual spherical coordinates $x^a=(\theta,\phi)$ on $S^2$.
Uppercase Latin indices $A,B,C,...$ denote objects on $M^2$ and
lowercase Latin indices $a,b,c,...$ denote objects on $S^2$.
The fact that we use arbitrary coordinates on $M^2$ will later allow
us to keep track of the tensorial character of the different variables.
Therefore we do not impose any condition on the lapse or shift,
apart from being consistent with spherical symmetry.

The 4-metric can be 2+2 decomposed as
\begin{equation}
(ds^2)_4 =
g_{AB}(x^D)\, dx^A dx^B + r^2(x^D)\, d\Omega^2,
\end{equation}
with $d\Omega^2$ the round metric of the 2-sphere and $g_{AB}$ and $r$
being a metric field and a scalar field on $M^2$, respectively. We define
the vector field $v_A \equiv r^{-1} r_{,A}$.

Using the radial coordinate $\rho$ explicitly we can write the background
spatial 3-metric as
\begin{equation}
(ds^2)_3 = a^2(t,\rho) d\rho^2 + r^2(t,\rho) d\Omega^2 .
\end{equation}
With a spherically symmetric lapse $\alpha=\alpha(t,\rho)$ and shift vector
$\beta^i=\left(\beta(t,\rho),0,0\right)$ we have
\begin{eqnarray}
(ds^2)_4 &=& ( - \alpha^2 + a^2\beta^2 ) dt^2 + 2 a^2 \beta dt d\rho + (ds^2)_3  \\
         &=& -\alpha^2 dt^2 + a^2(d\rho +\beta dt)^2 + r^2 d\Omega^2,
\end{eqnarray}
which takes the following matricial form,
\begin{equation}
g_{AB}=\left(
\matrix{ -\alpha^2+a^2\beta^2 & a^2\beta \cr
         a^2 \beta & a^2 }
\right) , \qquad
g^{AB}=\left(
\matrix{ -\alpha^{-2} & \alpha^{-2}\beta \cr
         \alpha^{-2} \beta & a^{-2}-\alpha^{-2}\beta^2 }
\right) .
\end{equation}
The normal vector to the surfaces of constant $t$ is
$u_\mu=(-\alpha,0,0,0)$ or $u^\mu=\alpha^{-1}(1,-\beta,0,0)$.
Its orthogonal, radial vector is $n^\mu=(0,a^{-1},0,0)$ or 
$n_\mu=a(\beta,1,0,0)$. In order to work with more geometrical
objects we define the following frame derivatives that act on any
scalar field $f$:
\begin{equation}
f'=n^\mu f_{,\mu}=\frac{f_{,\rho}}{a}, \qquad
\dot{f}=u^\mu f_{,\mu}=\frac{f_{,t}-\beta f_{,\rho}}{\alpha} .
\end{equation}

We now derive the background equations, that will be later used to
simplify the coefficients of the equations for the perturbations.
It is convenient to define the following momentum-like variables,
which have a definite tensorial character with respect to changes of the
$\rho$ coordinate,
\begin{equation}
\Pi_1 \equiv
\frac{a^2 \Pi^{\rho\rho}}{\mu_g} , \qquad 
\Pi_2 \equiv
\frac{2r^2 \Pi^{\theta\theta}}{\mu_g} , \qquad
\Pi_3 \equiv
\frac{\Pi}{\mu_g}.
\end{equation}
We can write the constraints in terms of these spherical variables,
\begin{eqnarray}\label{sphericalconstraint1}
\frac{\cal H}{\mu_g} &=&
\Pi_1\left(\frac{\Pi_1}{2}-\Pi_2\right)
- \three{R} +\frac{1}{2} \left({\Pi_3}^2+{\Phi'}^2\right)
= 0 ,
\\\label{sphericalconstraint2}
\frac{1}{a}\frac{{\cal H}_\rho}{\mu_g} &=&
-\frac{2}{r^2}(r^2\Pi_1)'+\frac{2r'}{r}\Pi_2
+ \Pi_3 \Phi' = 0 ,
\end{eqnarray}
so that the action is
\begin{equation}
\frac{1}{4\pi}{\cal S} = \int dt \int d\rho 
\; a r^2 \left[
  2\Pi_1\frac{a_{,t}}{a}
+ 2\Pi_2\frac{r_{,t}}{r}
+ \Pi_3 \Phi_{,t}
- \alpha \frac{\cal H}{\mu_g} 
- \beta \frac{{\cal H}_\rho}{\mu_g} \right] .
\end{equation}

The evolution equations can be obtained by simple variation with respect
to different variables:
\begin{eqnarray}
\frac{1}{\alpha}\left[a_{,t}-(\beta a)_{,\rho}\right] &=& \frac{a}{2} \left(\Pi_1-\Pi_2\right) , \\\label{Pi1}
\frac{1}{\alpha}(r_{,t}-\beta r_{,\rho}) &=& -\frac{r}{2} \Pi_1 , \\
\frac{1}{\alpha}(\Phi_{,t}-\beta \Phi_{,\rho}) &=& \Pi_3\\
\frac{1}{\alpha}\left( \Pi_1{}_{,t} -\beta \Pi_1{}_{,\rho } \right)
&=& 
\frac{3\Pi_1^2}{4} 
+ \frac{1}{r^2}
- \frac{r'}{r}\frac{(\alpha^2 r)'}{\alpha^2 r}
+ \frac{1}{4}\left(\Pi_3^2+{\Phi'}^2\right), \\
\frac{1}{\alpha}\left( \Pi_2{}_{,t} - \beta \Pi_2{}_{,\rho} \right)
&=&
\frac{1}{2}(\Pi_1^2+\Pi_2^2-\Pi_1\Pi_2)
+ \frac{2\alpha'r'}{\alpha r}
- \frac{2(\alpha r)''}{\alpha r}
+ \frac{1}{2}\left(\Pi_3^2-{\Phi'}^2\right), \\
\frac{1}{\alpha}(\Pi_3{}_{,t}-\beta \Pi_3{}_{,\rho}) &=& 
\frac{\Pi_3(\Pi_1+\Pi_2)}{2}
+\frac{(\alpha r^2\Phi')'}{\alpha r^2}.
\end{eqnarray}
Note that in spherical symmetry the restriction to vacuum, choosing Schwarzschild
coordinates ($t$, $r=\rho$), is given by $\Pi^{\mu\nu}=0$, $\three{R}=0$ and $\Phi=0$, that is,
$\Pi_1=\Pi_2=\Pi_3=\Phi=0$, which simplifies the previous expressions.
In particular, the constraints (\ref{sphericalconstraint1}) and
(\ref{sphericalconstraint2}) are then trivially obeyed. This is the case
studied by Moncrief \cite{Monc74}, but here we want to analyze the
general case.

\section{Axial perturbations}\label{section:axial}

\subsection{Harmonic expansions}
The tensor spherical harmonics form a complete basis on the 2-sphere to
expand a tensor field of any rank. Appendix \ref{appendix:tsh} gives
the definitions for the harmonics we shall need in this Article, as
well as some of their basic properties and the relations with the
harmonics used by Moncrief \cite{Monc74}. See Ref. \cite{BMM06}
for full definitions in the arbitrary rank case and further properties.
Tensor harmonics can be separated in two groups according to their
polarity:
there are {\em polar} (or {\em even}, {\em electric} or {\em poloidal})
harmonics and {\em axial} (or {\em odd}, {\em magnetic} or {\em toroidal})
harmonics. In first-order perturbation theory around a spherical spacetime
polar and axial harmonics decouple and this Article will only deal with
the axial part of the problem.
Following Regge-Wheeler's notation \cite{ReWh57} for the metric perturbations and Moncrief's
notations \cite{Monc74} for the momentum and shift vector, we expand the perturbative
variables in tensor spherical harmonics:
\begin{eqnarray}\label{expansion1}
h_{ij} dx^i dx^j &=& \sum_{l=1}^{\infty}\sum_{m=-l}^{l}\left\{
-2 (h{}_1)_l^m\, d\rho \; X_l^m{}_a\,dx^a + (h_2)_l^m \; X_l^m{}_{ab}\,dx^a dx^b\right\} , \\
\frac{1}{\mu_g} p_{ij} dx^i dx^j &=& \sum_{l=1}^{\infty}\sum_{m=-l}^{l}\left\{
-2 (\hat{p}_1)_l^m\, d\rho \; X_l^m{}_a\,dx^a + (\hat{p}_2)_l^m \; X_l^m{}_{ab}\,dx^a dx^b \right\}, \\
B_i dx^i &=& \sum_{l=1}^{\infty}\sum_{m=-l}^{l} - (h_0)_l^m \; X_l^m{}_a\, dx^a , \\
C &=& 0 , \\
p &=& 0 , \\\label{expansion6}
\varphi &=& 0 .
\end{eqnarray}
Note that there are no axial perturbations of the 3d scalars $\alpha$,
$\Phi$ and $\Pi$. That means that the scalar field plays no role from
the perturbative point of view, though the background scalar field is
still instrumental to allow for a general dynamical spacetime.
As we will see, this does not imply any loss of generality.
Different $(l,m)$ harmonic components also decouple around spherical
symmetry, and so from now on we shall drop them from the perturbative
variables, assuming that we work with a fixed pair of labels at any time.
It is important to note that $h_2$, $\hat{p}_2$ and $h_0$ are scalars
under changes of the $\rho$ coordinate, but $h_1$ and $\hat{p}_1$
behave as components of a vector. In the language of 1d spacetimes,
the latter are densities of weight +1, to be compensated with metric
factors $a$ to convert them into scalars.
This will become clearer when comparing with the more geometrical
GS approach.

The variables $(h_1, p_1)$ and $(h_2, p_2)$ form two pairs of canonically
conjugated variables, whose evolution is partially determined by the
arbitrary function $h_0$. For example the evolution equations for the
variables $h_1$ and $h_2$ can be easily obtained by perturbation of
formula (\ref{Pidef}) after introducing the expansions
 (\ref{expansion1}--\ref{expansion6})
\begin{eqnarray}
\frac{1}{\alpha}\left(h_{1,t} -(\beta h_1)_{,\rho}\right) &=&
2\hat{p}_1
+ \Pi_1 h_1
+\frac{r^2}{\alpha}\left(\frac{h_0}{r^2}\right)_{,\rho} , \\
\frac{1}{\alpha}\left(h_{2,t}-\beta h_{2,\rho}\right) &=& 
2\hat{p}_2
+(\Pi_2-\Pi_1)h_2
- \frac{2h_0}{\alpha} .
\end{eqnarray}
We shall later obtain the evolution equations for more convenient momenta
variables.

\subsection{Gauge-invariant perturbations}
The action functional for the axial perturbations is
\begin{eqnarray}
\frac{1}{2}\left(\delta^2{\cal S}\right)^{\rm axial} &=&
\int dt\int dx^3 \left[ p^{ij}h_{ij,t} -B^i \delta({\cal H}_i) + ... \right]^{\rm axial} \\\label{jaction}
&=& \int dt \left\{\int d\rho \; \left( p_1 h_{1,t} + p_2 h_{2,t} \right) + H[h_0] + ...\right\},
\end{eqnarray}
where the dots denote those terms coming from the second variation of
the constraints, which we do not need to consider in this subsection.
The functional $H$ will be defined below in terms of the first variation
of the constraint. We have also defined
\begin{equation}
p_1 = \frac{2l(l+1)}{a} \hat{p}_1^* ,
\qquad
p_2 = \frac{a\lambda}{r^2} \hat{p}_2^* ,
\end{equation}
where the star stands for complex conjugation and we have defined
\begin{equation}
\lambda \equiv \frac{1}{2}(l-1)l(l+1)(l+2) .
\end{equation}
In term of these variables, the perturbed constraint is given by
\begin{equation} \label{pert(H_a)axial}
\delta({\cal H}_a)^{\rm axial} =
\frac{X_a \mu_g}{l(l+1)} \left\{
 \frac{(r^2p_1)_{,\rho}}{ar^2}
+2\frac{p_2}{a}
+\lambda\frac{\Pi_2 h_2}{r^2}
+\frac{2l(l+1)}{ar^2}\left(\frac{r^2\Pi_1 h_1}{a}\right)_{,\rho}
\right\}
\end{equation}
which in turn defines the functional
\begin{eqnarray}
H[h_0] &=& - \int dx^3 B^i \delta({\cal H}_i)^{\rm axial}
\\
&=&
\int d\rho
\left\{
-r^2\left(\frac{h_0}{r^2}\right)_{,\rho}p_1
+2 h_0 p_2
+\lambda a\Pi_2\frac{h_0}{r^2}h_2
-\frac{2l(l+1)}{a}r^2\Pi_1 \left(\frac{h_0}{r^2}\right)_{,\rho}h_1
\right\} .
\end{eqnarray}

This functional is the generator of gauge transformations, and of
course commutes with itself on shell,
\begin{equation}
\left[\, H[f], H[g]\;  \right] = \frac{1}{l(l+1)}\int d\rho
\left\{r^4(f_{,\rho}g-g_{,\rho}f)\frac{1}{a}\frac{{\cal H}_\rho}{\mu_g}\right\},
\end{equation}
for arbitrary scalar fields $f$ and $g$.

Following Moncrief \cite{Monc74} we perform two canonical transformations
to separate the gauge-invariant information from the pure-gauge content
in the canonical pairs $(h_1,p_1)$ and $(h_2,p_2)$. The first canonical
transformation constructs the gauge-invariant combination $k_1$,
also a vector component,
\begin{equation}
k_1 \equiv h_1 + \frac{r^2}{2}\left(\frac{h_2}{r^2}\right)_{,\rho}
, \qquad
k_2 \equiv h_2 .
\end{equation}
It induces the following transformation on the momenta:
\begin{equation}
\pi_1 = p_1 , \qquad
\pi_2 = p_2 + \frac{(r^2 p_1)_{,\rho}}{2r^2} ,
\end{equation}
and can be obtained from the generating function
\begin{equation}
F(p_1,p_2,k_1,k_2)=p_1 k_1+ p_2 k_2 - p_1 \frac{r^2}{2}\left(\frac{k_2}{r^2}\right)_{,\rho}.
\end{equation}

In terms of the new variables we can write the first variation of the
axial constraint as:
\begin{equation}
\delta(H_a)^{\rm axial} =
\frac{X_a \mu_g}{l(l+1)} \left\{
2\frac{\pi_2}{a}
+\lambda\frac{\Pi_2 k_2}{r^2}
+\frac{2l(l+1)}{ar^2}\left[\frac{r^2\Pi_1}{a} \left(k_1 -\frac{r^2}{2}\left(\frac{k_2}{r^2}\right)_{,\rho}\right)\right]_{,\rho}
\right\} ,
\end{equation}
which does not contain $\pi_1$ and therefore commutes with
$k_1$. That is, $k_1$ is gauge-invariant, as we had anticipated.
This suggests the second canonical transformation:
\begin{equation}
Q_1 \equiv k_1, \qquad
Q_2 \equiv k_2 ,
\end{equation}
with conjugated momenta
\begin{eqnarray}
P_1 &\equiv& \pi_1 - l(l+1)\frac{r^2 \Pi_1}{a}\left(\frac{k_2}{r^2}\right)_{,\rho} , \\
P_2 &\equiv& \pi_2
+\frac{\lambda}{2r^2}a\Pi_2 k_2
+\frac{l(l+1)}{r^2}\left[\frac{r^2\Pi_1}{a} \left(k_1 -\frac{r^2}{2}\left(\frac{k_2}{r^2}\right)_{,\rho}\right)\right]_{,\rho} .
\end{eqnarray}
We can obtain this canonical transformation from the generating
function
\begin{equation}
F(P_1,P_2,k_1,k_2)=P_1 k_1+P_2k_2+
a \,l(l+1)\left\{
\frac{r^2\Pi_1}{a}\left(\frac{k_2}{r^2}\right)_{,\rho}\frac{k_1}{a}
-\Pi_1\left[\frac{r^2}{2a}\left(\frac{k_2}{r^2}\right)_{,\rho}\right]^2
-\frac{(l-1)(l+2)}{8r^2}\,\Pi_2k_2^2
\right\}.
\end{equation}
The first canonical transformation is independent of the dynamical
content of the background spacetime, in the sense that it does not
contain the background momenta $\Pi_1,\Pi_2,\Pi_3$. It is actually
identical to that of Moncrief \cite{Monc74}. For the sake of clarity,
we have separated the influence of the dynamical background into
the second canonical transformation, which trivializes for any static
background.

At this point we have isolated the physical information of the axial metric
perturbation in the pair $(Q_1,P_1)$ while the $(Q_2,P_2)$ contains 
the gauge subsystem. $P_2$ is the generator of gauge transformations,
\begin{equation}
\delta({\cal H}_a)^{\rm axial} = \frac{X_a \mu_g}{l(l+1)} \frac{2 P_2}{a}
\end{equation}
and hence it is gauge-invariant but constrained to vanish. Its conjugated
variable $Q_2$ is gauge-dependent and its time evolution is determined
by the arbitrary function $h_0$, which can be used to set any desired
value for $Q_{2,t}$.

\subsection{Evolution equations}
After replacing the new variables and integrating by parts a number of
times, we get the following Jacobi action:
\begin{equation}
\frac{1}{2}\left(\delta^2{\cal S}\right)^{\rm axial} =
\int dt \int d\rho \; 
\left[ P_1(Q_{1,t}-(\beta Q_1)_{,\rho})
     + P_2(Q_{2,t}-\beta Q_{2,\rho})
     +2 P_2 h_0
- \alpha {\cal H}^{(1)} \right],
\end{equation}
where we have defined the first-order quadratic Hamiltonian
\begin{eqnarray}
{\cal H}^{(1)}&\equiv& \Pi_1 (P_1 Q_1 - P_2 Q_2)
+ \frac{1}{a\lambda r^2}
  \left[ \left(\frac{r^2P_1}{2}+l(l+1)\frac{r^2\Pi_1}{a}Q_1\right)_{,\rho}-r^2P_2\right]^2 \\
 && + \frac{a P_1^2}{2l(l+1)}
+ \frac{l(l+1)}{a}\left[
\frac{(l-1)(l+2)}{2r^2}+\frac{\Pi_1(\Pi_2-\Pi_1)}{2}+\dot{\Pi}_1\right]
Q_1^2 . \nonumber
\end{eqnarray}

The variation of the action with respect to $h_0$ gives the constraint
that must be obeyed by the perturbations. This constraint now takes
the simple form
\begin{equation}
P_2=0 .
\end{equation}
This constraint is conserved in the evolution since variation with respect
to $Q_2$ gives
\begin{equation}
(r^2P_2)_{,t} = (\beta r^2 P_2)_{,\rho} .
\end{equation}
As $P_2$ is the generator of the gauge transformations, its conjugated
variable $Q_2$ is pure gauge. Its evolution
equation comes from taking the variation of the action respect to $P_2$,
\begin{equation}
\frac{1}{\alpha}(Q_2{}_{,t}-\beta Q_2{}_{,\rho}) = -2 \frac{h_0}{\alpha}
-\Pi_1 Q_2
-\frac{1}{a\lambda}\left(r^2P_1+l(l+1)\frac{2r^2\Pi_1}{a}Q_1\right)_{,\rho}.
\end{equation}
The initial data for $Q_2$ is gauge, and its evolution is fully determined
by the free function $h_0$. In particular it is possible to choose $Q_2=0$
initially and take $h_0$ so that $Q_2=0$ at all times.

We can obtain the physically relevant equations by variation of
the action with respect to the variables $(Q_1,P_1)$.
This gives rise to a system of two coupled second order equations in
$\rho$-derivatives, whose principal part is, in matricial form,
\begin{equation}\label{evolution}
\frac{(l-1)(l+2)}{2r^2}\;\;\frac{1}{\alpha}
\left(\begin{array}{c}\frac{2l(l+1)}{a}Q_1\\ P_1\end{array}\right)_{,t} \;=
\;\left(\begin{array}{cc} -\Pi_1 & -1 \\ \Pi_1^2 & \Pi_1\end{array}\right)
\;\; \frac{1}{a^2}
\left(\begin{array}{c}\frac{2l(l+1)}{a}Q_1\\ P_1\end{array}\right)_{,\rho\rho}
+ \dots
\end{equation}
the dots denoting lower order terms in $\rho$-derivatives of $Q_1$ and
$P_1$. We have divided $Q_1$ by $a$ to make it a scalar under changes of $\rho$ coordinate.
This is a second order in time evolution system, as corresponds to a single
wave-like degree of freedom, but it apparently has fourth order in
$\rho$-derivatives for generic values of the background variable $\Pi_1$.
This is false because the 2x2 matrix has always vanishing square, and hence
the system has third order at most. Actually it has second order, as can
be checked by taking the matrix to its Jordan canonical form.
Defining the combination
\begin{equation} \label{Ldef}
L \equiv P_1 + l(l+1)\frac{2\Pi_1}{a} Q_1
\end{equation}
the system (\ref{evolution}) is equivalent to the pair (we now use the dot and prime frame
derivatives to simplify the expressions):
\begin{eqnarray}
\label{Ldot}
(r^2 L)\dot{} &=& - 2\lambda \frac{Q_1}{a} , \\
\label{Qdot}
\left(-\frac{2\lambda}{r^2}\,\frac{Q_1}{a}\right)\dot{} &=&
\frac{1}{\alpha}\left(\frac{\alpha}{r^2}(r^2L)'\right)'
+\frac{\Pi_2-\Pi_1}{2r^2}(r^2 L)\dot{}
-\frac{(l-1)(l+2)}{r^2} L ,
\end{eqnarray}
which can be clearly combined into a single second order equation for $L$,
the sought generalization of the Regge-Wheeler equation for dynamical
backgrounds.

When restricting to vacuum the variable $rL/\lambda$ is the Cunningham-Price-Moncrief
master function \cite{CPM78} that obeys the Regge-Wheeler equation,
though it is not immediately related to the Regge-Wheeler
variable. Using the gauge $r=\rho, \beta=0$ in vacuum we have
$\Pi_1=0$ and hence $rL =r P_1$, while the Regge-Wheeler variable is
$Q_1/(a^2r)$. We have seen that the former is easily generalizable to
a dynamical situation as given in (\ref{Ldef}), but not the latter,
because it would require dividing by $\Pi_1$, which may vanish.

\subsection{The master scalar perturbation}
The physical variables $Q_1$ and $P_1$ are not scalars in
$M^2$ and therefore their values depend upon the foliation we have
chosen. It is better to describe the gravitational wave using not
only gauge-invariant information, but also foliation-invariant
information, that is, scalars in $M^2$.
Studying the geometric properties of those variables under changes of
foliation is not simple in the 3+1 notation. Following Gerlach
and Sengupta \cite{GeSe79}, we change to the $M^2$-adapted
framework introduced in Section \ref{section:bg}, in which the foliation
is described by the orthonormal frame $(u^A, n^B)$. This allows us
to look for scalars on $M^2$ as expressions which are frame independent.
The background momenta can be rewritten as
\begin{eqnarray}
\Pi_1 &=& -2 u^A v_A , \\
\Pi_2 &=& -2 u^A v_A -2 u^A{}_{|A} , \\
\Pi_3 &=& u^A \Phi_{,A} .
\end{eqnarray}
We see that $\Pi_1$ and $\Pi_3$ are essentially time components of
vectors in $M^2$. However $\Pi_2$ is a more complicated object.

In the GS formalism the axial part of the metric perturbation is 
decomposed in tensor spherical harmonics:
\begin{equation}\label{metricdecomposition}
\delta\left(g_{\mu\nu}\right) dx^\mu dx^\nu\equiv
h_{\mu\nu} dx^\mu dx^\nu\equiv
\sum_{l,m}
\left\{
(h_A)_l^m \; X_l^m{}_b dx^Adx^b+
{h}_l^m \; X_l^m\!{}_{ab} dx^adx^b
\right\}.
\end{equation}
For a given pair $(l,m)$ the vector $h_A$ and scalar $h$ are
related in the following way to the original Hamiltonian
variables (\ref{expansion1}),
\begin{eqnarray}
h_1 &=& -(h_\rho)_{GS} , \\
h_2 &=& 2 (h)_{GS} , \\
h_0 &=& \alpha^2(h^t)_{GS} .
\end{eqnarray}

The perturbations $h_A$ and $h$ are gauge-dependent, but the
following combination is gauge-invariant
\begin{equation}
\kappa_A \equiv h_A -r^2\left(\frac{h}{r^2}\right)_{,A} ,
\end{equation}
and fully contains the axial information. Therefore it must be given
in terms of the gauge-invariant variables $Q_1$, $P_1$ and $P_2$:
\begin{eqnarray}
\kappa_\rho &=& -Q_1 , \\
\kappa^t &=& 
\frac{1}{\alpha}\left[\hat{p}_2+\frac{\Pi_2}{2}h_2\right]
=\frac{1}{\lambda a\alpha}\left\{
r^2 P_2-\frac{1}{2}\left[r^2P_1+2l(l+1)\frac{r^2\Pi_1}{a}Q_1\right]_{,\rho}
\right\}.
\end{eqnarray}
Those relations can be inverted, giving
\begin{eqnarray}
Q_1 &=& -\kappa_\rho , \\
Q_2 &=& 2(h)_{GS} , \\
\frac{P_1}{l(l+1)} &=& 
-\epsilon^{AB}\kappa_{A,B}-2(n^Au^B+n^Bu^A) v_A \kappa_B , \\
\label{P2eq}
\frac{2}{a\alpha}\frac{r^2P_2}{l(l+1)} &=& (l-1)(l+2) \kappa^t
+\epsilon^{tC}\left\{r^4\epsilon^{AB}\left[r^{-2}\kappa_A\right]_{,B}\right\}_{,C}.
\end{eqnarray}
We see that the gauge-invariant $Q_1$ is the $\rho$ component of the
gauge-invariant vector $-\kappa_A$. Then $Q_2$ is a scalar in $M^2$,
but it is gauge-dependent. The momentum $P_1$ is the sum of two parts,
the first one being a scalar (the curl of the vector $\kappa_A$) and 
the second one being essentially the off-diagonal component of the 
symmetric tensor $v_{(A}\kappa_{B)}$.
Therefore $P_1$ is not a component of a tensor itself. Finally $P_2$ is,
apart from a factor $a\alpha$, the time component of a contravariant
vector. Again, it is important to stress that these properties are
very easy to obtain in GS formalism, but not in the original
Hamiltonian formalism, where the variables are well adapted to a
3d point of view.

We want to construct a scalar from a linear combination of $Q_1$ and
$P_1$, and perhaps their radial derivatives. We already know that
$P_1$ is the sum of a scalar plus a non-scalar, so that we have to
find whether it is possible to cancel that non-scalar part using $Q_1$.
It is possible because:
\begin{equation}
-2(n^Au^B+n^Bu^A)v_A\kappa_B = 
-2\epsilon^{AB}v_A\kappa_B - 4 (u^Av_A)(n^B\kappa_B) .
\end{equation}
The first term on the r.h.s. is a scalar and the second term is just
$-2\Pi_1 \frac{Q_1}{a}$.

Therefore the linear combination L (\ref{Ldef}) we defined in
the previous section is the scalar we were looking for. It is related
to the GS scalar variable as
\begin{equation}\label{master}
-\frac{1}{r^2}\frac{L}{l(l+1)}
=\Pi_{GS}\equiv \epsilon^{AB}\left[r^{-2}\kappa_A\right]_{,B}.
\end{equation}
This is the most important result of this paper: we have constructed
a gauge-invariant variable fully describing the physical content of
the axial gravitational wave and then we have shown that it is a scalar,
so that it is
also independent of the coordinate system used on the background
spacetime. Note that no other independent scalar can be formed as a
linear combination of $P_1$ and $Q_1$ and their derivatives.
For the cases when we $\Pi_1$ vanishes, which is equivalent to
the background gauge condition $\rho=r$ and $\beta=0$
[see Eq. (\ref{Pi1})], the variable $P_1$ is already a scalar.
Hence, it is again clear that $P_1$ is a more convenient variable
than $Q_1$ for these cases.

Gerlach and Sengupta showed \cite{GeSe79} that the master variable
(\ref{master}) obeys the following wave equation
\begin{equation} \label{GSaxial}
-\left[\frac{1}{2r^2}(r^4{\Pi}_{GS})^{|A}\right]_{|A}+\frac{(l-1)(l+2)}{2}{\Pi_{GS}}=
8 \pi\epsilon^{AB}\psi_{A|B},
\end{equation}
where the bar denotes the covariant derivative on the manifold $M^2$ and
$\psi_A$ is a gauge-invariant axial perturbation of the energy-momentum
tensor. This equation, equivalent to the pair (\ref{Ldot}--\ref{Qdot})
for scalar field matter,
is valid for all background spherical spacetimes and describes the
evolution of the axial gravitational wave coupled to any matter model.
Of course, different matter models will have additional variables and
equations coupled to (\ref{GSaxial}) but we stress the fact that both
the form of $\Pi_{GS}$ and the form of (\ref{GSaxial}) will remain
unchanged.

The only issue we still need to analyze is whether the
vector field $\kappa_A$, which could appear in the $\psi_A$ expression or
in those other equations for the matter variables, can always be
reconstructed from the $\Pi_{GS}$ scalar.
An important equation in the GS formalism is
\begin{equation}
(l-1)(l+2)\kappa_A = 16\pi r^2 \psi_A - \epsilon_{AB}(r^4\Pi_{GS})^{|B},
\end{equation}
whose time component gives (\ref{P2eq}), after imposing the
constraint $P_2=0$, and its radial component gives (\ref{Ldot}) for
scalar field matter. This equation can be solved algebraically for
$\kappa_A$ in terms of $\Pi_{GS}$ as long as $\psi_A$ does not
contain the symmetrized derivative $\kappa_{(A|B)}$ or higher derivatives
of $\kappa_A$. Second and higher derivatives of $\kappa_A$ can be ruled
out by requiring that the matter stress-energy momentum must not contain
second derivatives of the metric, because that would change the principal
part of the Einstein equations. Symmetrized first order derivatives of
$\kappa_A$ cannot be ruled out on physical grounds because perturbation
of covariant derivatives of tensor fields may introduce the term
\begin{equation} \label{badterm}
\delta^{axial}[\Gamma^a{}_{BC}] = X^a \, \frac{\kappa_{(B|C)}}{r^2}.
\end{equation}
This is the only possible source of symmetrized derivatives of $\kappa_A$;
all other perturbations of Christoffel give either $\Pi_{GS}$ or
undifferentiated $\kappa_A$ terms. However, we haven't found any standard
matter model in which (\ref{badterm}) appears under perturbation of its
stress-energy tensor.

The combination of Hamiltonian gauge methods with the imposition of
having a scalar field on $M^2$ has determined uniquely the Gerlach
and Sengupta scalar $\Pi_{GS}$, a variable containing all information
on the axial gravitational wave and obeying a master wave equation.
There are a number of ways of explaining the meaning of this
variable (see for instance \cite{Nolan}), but probably the simplest one
is given by the expression
\begin{equation}
\delta\left(R_{ABcd}\right) = - \frac{l(l+1)}{2}\; Y \;\epsilon_{AB}\;\epsilon_{cd} \; r^2 \Pi_{GS},
\end{equation}
where $Y$ is the scalar harmonic.

\section{Conclusions}\label{section:conclusions}
The use of linear perturbation theory avoids the intrinsic nonlinear
character of General Relativity, but introduces new problems that must
always be addressed in some form. A Hamiltonian approach to GR
perturbations allows a natural discrimination of the gauge from the
gauge-invariant information in the problem, but obscures the geometric
character of the perturbative variables. A complementary Lagrangian
approach offers a better geometric understanding, but gives no clues
on how to separate the physical, unconstrained content of the model
at hand. This Article proposes a combination of those two approaches
to arrive at a scalar, gauge-invariant and unconstrained description
of the linear perturbations in General Relativity.

Around a spherically symmetric background the axial and polar subsets
of perturbations decouple from each other; this Article has focused
on the axial subset, for which a metric master scalar was already
introduced by Gerlach and Sengupta \cite{GeSe79} using a purely
Lagrangian approach. Generalizing Moncrief's approach for a Schwarzschild
background \cite{Monc74}, we have reobtained this scalar for a general
time-dependent background. First, we have isolated the gauge-invariant
and unconstrained information in a Hamiltonian pair of
variables $(Q_1,P_1)$, each obeying a first order in time evolution equation. Then
we have analyzed the geometric character of these two variables, showing
that only a particular combination of them forms a scalar under
transformations on the reduced (under spherical symmetry) spacetime.
This scalar is, as was expected, the Gerlach and Sengupta master scalar.

The corresponding master scalar for the polar sector has never been
found for a general time-dependent background, and there is no known
obstruction for its existence. Such a variable would be relevant to
study, for example, the matching problem through a timelike surface
separating two different physical models (like fluid and vacuum at
a star surface \cite{MaGu01}). Therefore, the open question is whether
the same combination of techniques we have used in this article can be
successfully applied to the polar gravitational wave. This analysis
will be presented in a separate publication.

Many computations in this Article have been performed with the
{\em xAct} \cite{xAct, xPert, xPerm} framework for Tensor Computer Algebra. More precisely,
we have used the package {\em xPert} for metric perturbation theory around
general spacetimes to obtain the formulas of Section \ref{section:idea}.
Then, we have expanded these formulas in tensor spherical
harmonics with the package {\em Harmonics}.

\acknowledgments
We thank Guillermo A. Mena Marug\'an and Vincent Moncrief for discussions,
and the Albert Einstein Institute for hospitality.
DB acknowledges financial support from the FPI
program of the Regional Government of Madrid.
JMM acknowledges financial aid provided by the
I3P framework of CSIC and the European Social Fund.
This work was supported by the Spanish MEC Project
FIS2005-05736-C03-02.

\appendix
\section{Tensor spherical harmonics}
\label{appendix:tsh}
It is possible to construct harmonic bases for tensor fields of arbitrary
rank on the two-sphere $S^2$. This appendix briefly summarizes the
construction of those bases for scalars, vectors and symmetric tensors.
For the general case see \cite{BMM07}. Coordinates on $S^2$ will be
denoted with lowercase Latin letters $a,b,c,...$ and the round metric
will be $\gamma_{ab}$, with unit Gaussian curvature. Its associated
Levi-Civita connection will be denoted with a colon, such that
$\gamma_{ab:c}=0$. Finally, the volume form will be called $\epsilon_{ab}$.

The spherical harmonics $Y_l^m$ form a basis for scalar fields on $S^2$.
A vector basis can be constructed from them by differentiation,
following Regge and Wheeler, as $Z_a\equiv Y_{:a}$ and
$X_a\equiv \epsilon_{ab}\gamma^{bc}Y_{:c}$, the former being polar and
the latter axial.
Note that the labels $(l,m)$ are always implicitly assumed in the equations.
A basis for polar symmetric tensors is given by $\gamma_{ab} Y$
(pure trace) and $Z_{ab}\equiv Y_{:ab}+\frac{l(l+1)}{2}\gamma_{ab}Y$
(traceless).
Axial symmetric tensors can be expanded using the basis
$X_{ab}\equiv (X_{a:b}+X_{b:a})/2$.

These tensor harmonics are related to those of Moncrief \cite{Monc74}
as follows:
\begin{eqnarray}
(\hat{e}_1)_{ij} dx^i dx^j &=& - 2 d\rho \, X_a dx^a , 
\label{MoncriefGS_sa} \\
(\hat{e}_2)_{ij} dx^i dx^j &=& X_{ab} dx^a dx^b , \\
(\hat{f}_1)_{ij} dx^i dx^j &=& 2 d\rho \, Z_{a} dx^a, \\
(\hat{f}_2)_{ij} dx^i dx^j &=& d\rho^2 \, Y , \\
(\hat{f}_3)_{ij} dx^i dx^j &=& Y \gamma_{ab} \, dx^a dx^b , \\
(\hat{f}_4)_{ij} dx^i dx^j &=& Y_{a:b} \, dx^a dx^b .
\end{eqnarray}
Under integration over the two-sphere the tensor spherical harmonics
form an orthogonal basis, with the following normalizations:
\begin{eqnarray}
\int_{S^2} d\Omega \; Y^* \; Y' &=& \delta_{ll'} \delta_{mm'} , \\
\int_{S^2} d\Omega \; \gamma^{ab} \; Z^*_{a} \; Z'_{b} &=& l(l+1) \delta_{ll'} \delta_{mm'} , \\
\int_{S^2} d\Omega \; \gamma^{ab} \; X^*_{a} \; X'_{b} &=& l(l+1) \delta_{ll'} \delta_{mm'} , \\
\int_{S^2} d\Omega \; \gamma^{ab} \gamma^{cd} \; Z^*_{ac} \; Z'_{bd} &=& \frac{1}{2}\frac{(l+2)!}{(l-2)!} \delta_{ll'} \delta_{mm'} , \\
\int_{S^2} d\Omega \; \gamma^{ab} \gamma^{cd} \; X^*_{ac} \; X'_{bd} &=& \frac{1}{2}\frac{(l+2)!}{(l-2)!} \delta_{ll'} \delta_{mm'} .
\end{eqnarray}




\end{document}